\documentstyle[12pt]{article}
\begin{document}
\title{\vspace*{-1.8cm}
Inelastic Rescattering in $B$ Decays to
$\pi \pi$, $\pi K$, and $K\bar{K}$,
and Extraction of $\gamma $}
\author{
{P. \.Zenczykowski}$^*$\\
\\
{\em Dept. of Theoretical Physics},
{\em Institute of Nuclear Physics}\\
{\em Radzikowskiego 152,
31-342 Krak\'ow, Poland}\\
}
\maketitle
\begin{abstract}
We discuss all contributions from inelastic
SU(3)-symmetric  rescattering 
in $B$ decays into a final pair of pseudoscalar mesons 
$PP = \pi \pi$, $K\bar{K}$, $\pi K$. FSI-induced
modifications of amplitudes obtained from the quark-line
approach are described in terms of a few parameters which take care of
all possible SU(3)-symmetric forms relevant for final-state
interactions.
Although in general it appears impossible to uniquely determine FSI effects
from the combined set of all $\pi \pi$, $K \bar{K}$, and $\pi K$ data,  
drawing some conclusions is feasible.
In particular, it is shown that in leading order 
the amplitudes of strangeness-changing $B$ decays 
depend on only one additional complex FSI-related parameter
apart from those present in the definitions of penguin
and tree amplitudes. 
It is also shown that joint considerations of 
U-spin-related
$\Delta S =0$ and $|\Delta S|=1$ decay amplitudes are modified
when non-negligible SU(3)-symmetric FSI are present.
In particular, if rescattering in $B^+ \to K^+\bar{K}^0 $
is substantial,
determination of the CP-violating
weak angle $\gamma $
from $B^+ \to \pi ^+ K^0$, $B^0_d \to \pi^-K^+$,  $B^0_s \to \pi ^+ K^-$,
and their CP counterparts
might be susceptible to important FSI-induced corrections.
\end{abstract}
$^*$ E-mail:
zenczyko@iblis.ifj.edu.pl
\newpage

\section{Introduction}
Most of the literature analysing 
CP-violating effects in $B$ decays
(with $B \to PP =\pi \pi$, $K\bar{K}$, $\pi K$ in particular)
deals with quark-diagram
short-distance (SD)
amplitudes and assumes that final state interactions (FSI) are negligible.
On the other hand, it has been argued that this neglect
is not justified and that any reliable analysis must take FSI into
account \cite{Wolf91,Don96etal,Zen97}. 
Indeed, recent analyses seem to
show that even in 
$B \to D^* X $ 
decays FSI must
play an important role (see eg. \cite{Cheng01}).
Accordingly,
various authors have tried to estimate FSI in $B \to PP$ decays
by analysing the contribution from
elastic or quasi-elastic rescattering \cite{Zen99}. 
The main problem, however,
 is posed by the sequence
$B \stackrel{weak}{\rightarrow} i \stackrel{FSI}{\rightarrow} PP $ 
involving {\em inelastic} rescattering processes
$i \stackrel{FSI}{\rightarrow} PP$, where $i$ denote all kinds of
multiparticle states. 
Arguments have been given that these inelastic processes  
constitute the main source of soft FSI phases 
\cite{Wolf91,SuzWolf99}. Since
estimates of the size of these effects are model-dependent,
one may envisage various scenarios, with the contributions from different 
intermediate states cancelling in an approximate way or renormalizing
SD prescriptions without changing their form, 
having random phases \cite{SuzWolf99}, or adding coherently \cite{Zen2001},
just to mention a few possibilities.
With our insufficient
knowledge of $PP$ interactions at $\sqrt{s} =m_B \approx 5.2~GeV$,
there is virtually no hope 
that a reliable calculation of inelastic FSI can be performed.

Consequently, various authors have argued that perhaps one should try to
determine FSI effects directly from the data. 
For example, decays $B^o_d \to K^+ K^-$
are thought to provide a measure on the size of FSI effects \cite{FSIK+K-}.
With many different decay channels and three  varieties of $B$ mesons 
($B^+$, $B^o_d$, $B^o_s$) one may hope that the FSI effects 
can be untangled,
especially if simple SU(3)-symmetric FSI is accepted.
As FSI are oblivious of the original decay mechanism,
various decays  (for example, independently of whether
the decay is strangeness-conserving or changing)
are affected by the same SU(3)-symmetric FSI. 
If these FSI can be described with the help of a few parameters only,
one may hope that
the number of measurable decay types might be sufficient to
permit determination of these parameters.
Learning the size of FSI directly from the data would be certainly
important as there are various papers which fit the present data on $B \to
\pi \pi, \pi K, K \bar{K}$ decays both without 
and with FSI (eg. \cite{HouYang}, \cite{ZWNG}).

The SD approaches attempt to include all strong interaction effects
by assigning different phase parameters to different quark-line diagrams
(eg. tree $T$, penguin $P$, etc.).
However, it was argued that this prescription violates 
 such tenets of strong interactions as isospin symmetry \cite{Wolf95,GW97}. 
The origin of the problem pointed out in ref. \cite{GW97} 
is the lack of any (isospin) correlation between the spectator quark and the
products of $b$ quark decay. 
By its very nature such correlation cannot be provided by SD
dynamics. A long-distance (LD) mechanism
which ensures that quarks "know" about each other must be involved here. 
The inelastic
rescattering effects considered in the present paper will provide
both such a correlation and a generalization of the formulas
of ref.\cite{GW97}.
We shall show how the standard formulas of the SD approach to $B$ decay
amplitudes are modified when FSI are not negligible.
In particular, assuming the dominance of SD dynamics by a few (2 or 3) 
quark-line amplitudes (as it is usually done) we will discuss ways
in which deviations from these formulas can be used to indicate
the size of inelastic FSI (IFSI).
It will be also shown that rescattering may affect considerations
based on analyses of U-spin related decays, including the
method of extracting the value of the CP-violating weak angle $\gamma $
from $B \to \pi K$ decays.

\section{General}
If one accepts that final state interactions cannot modify the probability of
the original SD weak decay, it follows that vector $\bf W$ representing
the set of all FSI-corrected amplitudes is related to vector $\bf w$ of
the original amplitudes driven by the SD dynamics through \cite{Zen2001}:
\begin{equation}
\label{S12}
{\bf W}={\bf S}^{1/2}{\bf w}\approx (1+\frac{1}{2}({\bf S}-1)+...){\bf w}.
\end{equation}
After the SD-driven $B \to PP$ decay whose description is included in 
$\bf w$, 
the  $PP$ pair produced may undergo
further scattering into many non-$PP$ states. 
This out-of-$PP $-channel  process provides absorption in the $PP$ channel,
i.e. it reduces the original decay amplitudes.
This is
described by (mainly
imaginary) Pomeron exchange contribution in ${\bf T} $ 
(${\bf S}-1 = i{\bf T} \to - {\rm Im}~T $).

Pomeron contributions in direct channels belonging
to different SU(3) multiplets are related using 
$u-d-s$ symmetry of the quark diagram approach. 
This approach relates 
absolute magnitudes and phases of FSI
amplitudes in various direct channels
corresponding to different SU(3) multiplets.
(SU(3) itself, on the other hand, relates amplitudes only within - 
but not between - these channels.)
For Pomeron, the
 FSI effects 
in all possible SU(3) channels (${\bf 1}$, ${\bf 8}$, ${\bf 27}$)
are identical.
Thus, Pomeron exchange between departing pseudoscalar mesons
amounts to rescaling down the overall size of all quark-line 
decay amplitudes without
modifying any other SD predictions.  

The $b \to u\bar{u}q$  and $b \to c\bar{c}q$ SD decay processes
lead directly also to non-$PP$ 
states composed of two higher-mass states (resonances) $M_1$ and $M_2 $.
The latter
may rescatter into $PP$ yielding an "indirect" contribution to
$B \to PP$.
Thus, the set of FSI-corrected decay amplitudes ${\bf W} =[ W_j ] $ 
is composed of
the direct and indirect parts as follows
(amplitudes $w_i$ are already absorption-rescaled):
\begin{equation}
\label{dirplusresc}
W_j = w_j + \sum_{k,\alpha} F_{j,k \alpha }w_{k \alpha}
\end{equation}
where the indirect contributions are described by the sum on the r.h.s.
In Eq.(\ref{dirplusresc})
the subscripts  denote  decay channels rather schematically:
$j,k$ are SU(3)-related indices, while $\alpha $ labels inelastic channels.
In this paper we are interested in finding the pattern of inelastic
FSI contributions
following the original SD decay $b \to u\bar{u}q$. Rescattering from the
$b \to c\bar{c}q$-generated intermediate states leads to charming
penguins \cite{Martinelli}, whose amplitudes may be added to those of
SD penguins in the final formulas.

Formally, the choice of decay channels $j$ (i.e. a basis in the 
flavour space) 
is irrelevant, and one may
use either a Cartesian basis (where all mesons in $PP$ states have definite
$q\bar{q}$ content), or SU(3) basis (in which $j$ correspond to
- belonging to different SU(3) multiplets -
linear combinations of $(q\bar{q})(q\bar{q})$).
However, as resonances appear only in the octet
channel, FSI in the octet and the 27-plet channels are different.
Consequently, 
it is natural to use the SU(3) basis,  only at the end
transforming everything to the basis of interest.

Consider now the simple case when SU(3) is replaced by SU(2)
and $j,k = {\bf 1}, {\bf 3}, {\bf 5}, ...$ label SU(2) multiplets.
Furthermore, in order to simplify the argument,
let us assume that for all $\alpha =1,...N$ one has
$w_{k \alpha } = w_k$  and 
$F_{j,k\alpha }=F_{j,k}$.
Clearly, we must have $F_{j,k}=f_j\delta _{jk}$
with $f_j$ complex in general.
One obtains then 
\begin{equation}
\label{simplediagbreak}
W_j=(1+Nf_j)w_j.
\end{equation}
If $f_j=f$ for all $j$,
one has ${\bf W}= (1+Nf){\bf w}$, i.e. all FSI-induced modifications
are contained in one, experimentally not discernible,
 overall complex factor $1+Nf$, identical for all isopin channels.
If strong interactions in different isospin channels are different
(i.e. $f_j\ne f_i$ for $j \ne i$), 
the differences between $f_j$'s will
lead to a modification of the SD pattern: the
magnitudes and phases of FSI effects will depend on isospin.

One expects the SU(3) case to be similar:
for an appropriate choice of $F$'s in Eq.(\ref{dirplusresc}),
no FSI should be discernible in the final $W_j$ amplitudes.
Modifications of the predictions of the 
SD quark-line approach may appear only when
 FSI in different SU(3) channels differ from this particular choice.
The relevant conditions on the SU(3) analogues of $f_i$ 
 are derived in Section 4.

\section{SD amplitudes for decays into inelastic SU(3) eigenstates}
In this paper we accept SU(3) in both direct and indirect
terms as we do not attempt to fit any data as yet.
When doing the latter, 
SU(3) breaking should probably be first introduced  in the direct term,
as one may argue that no corrections to corrections 
(i.e. no SU(3)-breaking in FSI effects) should be considered
in the first attempt.

Our conventions and definitions for the (final, symmetrized) $PP$ states 
are given 
in the Appendix, where $PP$ states with mesons of definite charges, 
$PP$ states of definite isospin, and $PP$ states belonging to definite SU(3)
multiplets (i.e. direct-channel SU(3) eigenstates) are listed. 

In quasi-elastic FSI the intermediate state is also a $PP$ state, and thus
the intermediate mesons have to be symmetrized. In the inelastic case
the original SD weak decay produces two $q\bar{q}$ pairs, which transform
into a pair of resonances $M_1M_2$.
These $M_1$ and $M_2$ mesons
are different in general 
(we neglect the case when the two mesons are identical as
the bulk of inelastic rescattering must come from $M_1 \ne M_2$).
We may define $M_1$ to be the state of lower mass.
In the Appendix we call the first (second) meson $M_1$ ($M_2$)
a $P$ ($V$) meson.
Here $P$ and $V$ are only labels
denoting different SU(3) multiplets of mesons, 
such as pseudoscalar, vector, 
axial, tensor etc. (including heavier and heavier) mesons. 
With $P \ne V$, there is no need to symmetrize.
In particular, the $PV$
states do not have to be symmetric in SU(3) indices.
Thus, while in the case of quasi-elastic FSI the mesons $V$ and $P$ 
are both pseudoscalars
and only states $(P_aP_b+P_bP_a)/\sqrt{2}$ (with  $P$
representing a pseudoscalar
and $a$, $b$
being SU(3) indices)
are admissible, in general we must distinguish
cases when $M_1M_2=P_aV_b$ and $M_1M_2=P_bV_a$ .  
Using the $PV$ labels to
denote all such situations, the Appendix lists
all the relevant $PV$ states in the SU(3) basis.
In the preparation of this list one has to consider
both SU(3)-symmetric and SU(3)-antisymmetric combinations of 
octet mesons $P$ and
$V$ in particular. In order
to prevent any misunderstanding, we note that the replacement
$P \rightleftharpoons V$ has nothing to do with this
SU(3) (anti)symmetrization:
indices $P$, $V$ do not belong to the SU(3) group as is explicit
in the Appendix.
Note that while the ${\bf 27}$-plet can be obtained 
only in the ${\bf 8} \times
{\bf 8}$ $PV$ channel, the octet may be obtained not only as a
 symmetric or antisymmetric combination of two octets, but also
 from a singlet $P$ and octet $V$ (or vice versa). Similar
 possibilities exist for the singlet $PV$ channel.
 Since in each of these $PV$ channels  
 ($({\bf 8} \times {\bf 8})\to {\bf 27}, {\bf 8}_s, {\bf 8}_a, {\bf 1}$; 
 (${\bf 8}
 \times {\bf 1}) \to {\bf 8}$,  etc. ) 
 rescattering of generally unknown
 form may take place, one is forced to use a free parameter 
 to describe FSI in each such given channel. This proliferation of free
 parameters constitutes the main obstacle on the way of their 
 determination from data.

Possible types of SD diagrams are shown in Fig. 1. 
For $T$ (tree), $C$ (colour-suppressed), $P$ (penguin), $S$
(singlet penguin) amplitudes
only these diagrams 
are shown in which short-distance $b$ decay
consists in the emission of meson $M_1=P$
 off the decaying quark line  
(i.e. when the spectator quark is
not taken into account).
These amplitudes are denoted 
by  $T_1$, $P_1$, $C_1$, ... 
for strangeness-conserving processes
($T_1'$, $P_1'$, $C_1'$,... for strangeness-changing processes).
When short-distance
$b$ decay produces meson $M_2=V$, 
the corresponding amplitudes (not shown in Fig.1) are denoted 
by $T_2$, $P_2$ etc.
($T_1$ does not have to be equal to $T_2$).
Although 
we keep the distinction between $E_1$ and $E_2$ 
as well as $A_1$ and $A_2$, 
 in these cases
quarks produced in
$\bar{b}d (\bar{b}s)$ should enter $P$ and $V$ mesons
with equal probabilities. 
For the penguin annihilation 
amplitudes ($PA$ and $SS$)
there does not seem to be any reason 
why  $PA_1 \ne PA_2$ or $SS_1 \ne SS_2$, 
hence $PA$ and $SS$ do not carry a subscript.

With the above preparations,
 the amplitudes for strangeness-conserving $\Delta S=0$ 
 (strangeness-violating $\Delta S =1$) decays
 into quasi-two-body "$M_1M_2$" SU(3) channels 
 may be calculated 
 in terms of unprimed (primed)
 SD quark-line amplitudes $T_i$, $P_i$, ... ($T'_i$, $P'_i$, ...).
 We label channels by their SU(3) and isospin characteristics, e.g.
 $({\bf 8}_a,1)$ denotes an isospin-1 octet channel formed as an
 antisymmetric combination of $P_8$ and $V_8$.
 
 With the channels being specified on the l.h.s. 
 and denoting $T_1+T_2=2T$, $P_1+P_2=2P$, $C_1+C_2=2C$, 
 $A_1+A_2=2A$, $E_1+E_2=2E$, and similarly for the primed
 amplitudes,
 one obtains the following
 expressions

a) for $B^+$ decays
\begin{eqnarray}
({\bf 27},2)      && -(T+C)\nonumber \\
({\bf 27},3/2)    &&\frac{2}{\sqrt{6}}(T'+C')\nonumber\\
({\bf 27},1)      &&-\frac{1}{\sqrt{5}}(T+C)\nonumber\\
({\bf 27},1/2)    &&2\sqrt{\frac{2}{15}}(T'+C')\nonumber\\
({\bf 8}_s,1)     &&-\frac{2}{\sqrt{30}}(T+C+5 P
+5 A)\nonumber\\
({\bf 8}_s,1/2)   &&\frac{2}{\sqrt{30}}(T'+C'+5 P'
+5 A')\nonumber\\
({\bf 8}_a,1)     &&-\frac{2}{\sqrt{6}}(T-C+3 P
+3 A)\nonumber\\
({\bf 8}_a,1/2)   &&\frac{2}{\sqrt{6}}(T'-C'+3 P'
+3 A')\nonumber\\
({\bf 8}_{81},1)  &&-\frac{1}{\sqrt{3}}(T_1+C_2+2P+2A)\nonumber\\
({\bf 8}_{81},1/2)&&\frac{1}{\sqrt{3}}(T'_1+C'_2+2P'+2A')
\nonumber\\
({\bf 8}_{18},1)  &&-\frac{1}{\sqrt{3}}(T_2+C_1+2P+2A)\nonumber\\
\label{eq4}
({\bf 8}_{18},1/2)&&\frac{1}{\sqrt{3}}(T'_2+C'_1+2P'+2A')
\end{eqnarray}
b) for $B^0_d$ decays
\begin{eqnarray}
({\bf 27},2)      &&-\frac{2}{\sqrt{6}}(T+C)\nonumber\\
({\bf 27},3/2)    &&\frac{2}{\sqrt{6}}(T'+C')\nonumber\\
({\bf 27},1)      &&0\nonumber\\
({\bf 27},1/2)    &&\frac{2}{\sqrt{30}}(T'+C')\nonumber\\
({\bf 27},0)      &&-\frac{1}{\sqrt{30}}(T+C)\nonumber\\
({\bf 8}_s,1)     &&\sqrt{\frac{5}{3}}(E-P)\nonumber\\
({\bf 8}_s,1/2)   &&\frac{2}{\sqrt{30}}(3T'-2C'+5P')
\nonumber\\
({\bf 8}_s,0)     &&-\frac{2}{3\sqrt{20}}(6T-4C+5P
+5E)\nonumber\\
({\bf 8}_a,1)     &&-\frac{1}{\sqrt{3}}(2T+3P-3E)
\nonumber\\
({\bf 8}_a,1/2)   &&\frac{2}{\sqrt{6}}(T'+3P')\nonumber\\
({\bf 8}_a,0)     &&-(E+P)\nonumber\\
({\bf 8}_{81},1)  &&\frac{1}{\sqrt{6}}(C_1-C_2-2P+2E-S_2)
\nonumber\\
({\bf 8}_{81},1/2)&&\frac{1}{\sqrt{3}}(C'_2+2P'+S'_2)\nonumber\\
({\bf 8}_{81},0)  && -\frac{1}{3\sqrt{2}}(2C+2P+2E+S_2)
\nonumber\\
({\bf 8}_{18},1)  &&\frac{1}{\sqrt{6}}(-C_1+C_2-2P+2E-S_1)
\nonumber\\
({\bf 8}_{18},1/2)&&\frac{1}{\sqrt{3}}(C'_1+2P'+S'_1)\nonumber\\
({\bf 8}_{18},0)  &&-\frac{1}{3\sqrt{2}}(2C+2P+2E+S_1)
\nonumber\\
({\bf 1}_{88},0)  &&\frac{1}{3\sqrt{2}}(3T-C
+8P+8E+12 PA)\nonumber\\
({\bf 1}_{11},0)  &&\frac{1}{3}(2C+2P+2E+3 PA
+2S +SS) \nonumber\\
&&
\end{eqnarray}
c) for $B^0_s$ decays
\begin{eqnarray}
({\bf 27},2)      &&0\nonumber\\
({\bf 27},3/2)    &&-\frac{2}{\sqrt{6}}(T+C)\nonumber\\
({\bf 27},1)      &&-\frac{2}{\sqrt{10}}(T'+C')\nonumber\\
({\bf 27},1/2)    &&-\frac{2}{\sqrt{30}}(T+C)\nonumber\\
({\bf 27},0)      &&\sqrt{\frac{3}{10}}
(T'+C')\nonumber\\
({\bf 8}_s,1)     &&\frac{1}{\sqrt{15}}(3T'+5E'
-2C')\nonumber\\
({\bf 8}_s,1/2)   &&-\frac{2}{\sqrt{30}}(3T-2C+5P)
\nonumber\\
({\bf 8}_s,0)     &&\frac{1}{3\sqrt{5}}(3T'-2C'
+10P'-5E')\nonumber\\
({\bf 8}_a,1)     &&-\frac{1}{\sqrt{3}}(T'-3E') \nonumber\\
({\bf 8}_a,1/2)   &&-\frac{2}{\sqrt{6}}(T+3P) \nonumber\\
({\bf 8}_a,0)     &&(T'+2P'-E')
\nonumber\\
({\bf 8}_{81},1)  &&\frac{1}{\sqrt{6}}(C'_1+2E')\nonumber\\
({\bf 8}_{81},1/2)&&-\frac{1}{\sqrt{3}}(C_2+2P)\nonumber\\
({\bf 8}_{81},0)  &&-\frac{1}{3\sqrt{2}}(C'_1-2C'_2-4P'+2E')
\nonumber\\
({\bf 8}_{18},1)  &&\frac{1}{\sqrt{6}}(C'_2+2E')\nonumber\\
({\bf 8}_{18},1/2)&&-\frac{1}{\sqrt{3}}(C_1+2P)\nonumber\\
({\bf 8}_{18},0)  &&-\frac{1}{3\sqrt{2}}
(C'_2-2C'_1-4P'+2E')\nonumber\\
({\bf 1}_{88},0)  &&\frac{1}{3\sqrt{2}}(3T'-C'+8P'
+8E'+12 PA')\nonumber\\
({\bf 1}_{11},0)  &&\frac{1}{3}(2C'+2P'+2E'+3PA'+
2S'+SS').\nonumber\\
\label{eq6}
&&
\end{eqnarray}

\section{Modifications of SD amplitudes due to inelastic rescattering}

Usually, the SD quark-diagram analyses of $B \to PP$ decays start
with an assumption that only two or three diagrams types
are dominant, while
the remaining ones are negligible.
Thus, in strangeness-conserving  ($b \to ud\bar{u}$) decays one expects the
hierarchy $|T| > |P|, |C| > ...$ \cite{GHLR95}, while in the
strangeness-violating decays one expects $|P'| > |T'|> ...$.
Denoting the amplitudes for decays into a given $M_1M_2$ state
with superscript $^{(\alpha )}$, we substitute in Eqs.(\ref{eq4}-\ref{eq6})
$T \to T^{(\alpha )}$, $P \to P^{(\alpha )}$, etc.
Since at the level of short-distance decay it is not yet decided  
whether the particular
quark-level state will hadronize as the $PP$ state
or one of the $M_1M_2$ states,  
one expects that quark-level amplitudes for the $B \to M_1M_2$ 
and $B \to PP$
transitions exhibit the same hierarchy pattern.
Thus, transition amplitudes
$T^{(\alpha )}$, $C^{(\alpha )}$, $P^{(\alpha )}$  should
satisfy $T^{(\alpha )}=\eta^{(\alpha )}T>C^{(\alpha )}=
\eta ^{(\alpha )}C,P^{(\alpha )}=\eta ^{(\alpha )}P>...$
with $T$, $C$, $P$ now describing transitions into pseudoscalar pairs, 
and analogously for primed amplitudes
($\eta _a$ takes care of an overlap between quark-level and hadron-level 
states).

We will consider IFSI corrections 
resulting from the inelastic rescattering of
the $M_1M_2$ states generated by these dominant amplitudes 
($T^{(\alpha )},P^{(\alpha )},C^{(\alpha )}$) and 
($P'^{(\alpha )},T'^{(\alpha )}$) into $PP$. 
We will not keep any other terms, even though there are known
problems with the description of $B \to \eta ,\eta '$ decays, which
indicate that in these decays
the contributions from singlet penguin amplitudes 
may be significant.
One expects, however, that 
contributions in which intermediate states
are generated by Zweig-rule-violating SD amplitudes
should be negligible for general (non-$PP$) inelastic states.

We describe inelastic final state interactions by introducing several
free parameters as follows:
\begin{eqnarray}
(M_1({\bf 8})M_2({\bf 8}))_{\bf 27} \to (PP)_{\bf 27}&
& f^{(\alpha )}_{27}\nonumber \\
(M_1({\bf 8})M_2({\bf 8}))_{{\bf 8}_s} \to (PP)_{\bf 8}&
& f^{(\alpha )}_{s}\nonumber \\
(M_1({\bf 8})M_2({\bf 8}))_{{\bf 8}_a} \to (PP)_{\bf 8}&
& f^{(\alpha )}_{a}\nonumber \\
M_1({\bf 1})M_2({\bf 8}) \to (PP)_{\bf 8}&
& f^{(\alpha )}_{1,8}\nonumber \\
M_1({\bf 8})M_2({\bf 1}) \to (PP)_{\bf 8}&
& f^{(\alpha )}_{8,1} \nonumber \\
M_1({\bf 8})M_2({\bf 8}) \to (PP)_{\bf 1}&
& f^{(\alpha )}_{8,8}\nonumber \\
M_1({\bf 1})M_2({\bf 1}) \to (PP)_{\bf 1}&
& f^{(\alpha )}_{1,1}
\end{eqnarray}

Upper indices label inelastic intermediate states in the direct channel
 (some 
$f^{(\alpha )}$ 
may be zero).

Let us now consider as an example
the $B^+$ decay into the ${\bf 27}$-plet $PP$ state.
One calculates that (with the direct term already 
including absorption-induced
rescaling)
\begin{equation}
\label{sumexample}
W(B^+\to PP({\bf 27},1))=-\frac{1}{\sqrt{10}}(T+C)-
\frac{1}{\sqrt{10}}
\sum_{\alpha }f^{(\alpha )}_{27}(T^{(\alpha )}+C^{(\alpha )}).
\end{equation}
Using $T^{(\alpha )}=\eta ^{(\alpha )} T$ etc.,
the above equation may be reduced to
\begin{equation}
\label{sum}
W(B^+\to PP({\bf 27},1))=-\frac{1}{\sqrt{10}}(T+C)(1+f_{27})
\end{equation}
where $f_{27}\equiv \sum _{\alpha }f^{(\alpha )}_{27}\eta ^{(\alpha )}$.
We observe that the original amplitude has been multiplied by
an inessential complex factor $1+f_{27}$, which may be absorbed into
the definition of $T$ and $C$.

Following the above example, one introduces parameters
$f_s$, $f_a$, $f_{1,8}$, $f_{8,1}$, $f_{8,8}$ and $f_{1,1}$.
As these parameters are free, in order to keep the
formulas simple we define some of the parameters with additional
purely numerical factors included.
One may expect  that $P_1V_8 \to (PP)_8$
and $P_8V_1 \to (PP)_8$ are roughly similar.
Thus, FSI-induced contributions proportional
to the difference $f_{1,8}-f_{8,1}$ should be smaller 
than those proportional to $f_{1,8}+f_{8,1}$. 
Consequently, for simplicity we shall assume that 
$f_{1,8}=f_{8,1}$.

Proceeding as in the example leading to Eq.(\ref{sum}),
we may derive (after transforming to the basis in which final mesons
are in states of
definite charge):
\begin{eqnarray}
W(B^+\to\pi ^+ \pi ^0) &=&
-\frac{1}{\sqrt{2}}(T+C)(1+f_{27})\nonumber \\
W(B^+\to K^+ \bar{K}^0) &=& -P(1+f_{27})\nonumber\\&&
-\frac{1}{5}\{ T\Delta _1 +
P\Delta _2 + C\Delta _3 \} 
\end{eqnarray}
where
\begin{eqnarray}
\Delta _1&=&(f_s-f_{27})+f_a+f_{1,8} \nonumber \\
\Delta _2&=&5(f_s-f_{27})+3f_a+2f_{1,8} \nonumber\\
\label{eq11}
\Delta _3&=&(f_s-f_{27})-f_a+f_{1,8}. 
\end{eqnarray}
The above equations reduce to standard SD prescriptions (with an overall
factor of $1+f_{27}$) when $\Delta _1 = \Delta _2 = \Delta _3 =0$, i.e.
when $f_s-f_{27}=f_a=f_{1,8}=0$. This is the explicit form of the
condition for no observable FSI effect, mentioned in Section 2.

Having presented the general idea,
we now list all the relevant formulas.
The decays in which at least one pseudoscalar produced is  $\eta $ or
$\eta '$ involve additional uncertainties at the direct level. Consequently,
using these decays to help untangle the FSI is risky. Thus,
we restrict ourselves to $B$ decays into $\pi \pi$, $\pi K (\bar{K})$, 
and $K\bar{K}$.

In the $\Delta S =0$ sector, keeping only the $T$, $P$, $C$ terms, we have
\begin{eqnarray}
W(B^+\to \pi ^+ \pi^0) &= &-\frac{1}{\sqrt{2}}(T+C)(1+f_{27})\nonumber \\
W(B^+\to K^+ \bar{K}^0) &=& -P(1+f_{27})\nonumber\\&&
-\frac{1}{5}\{ T\Delta _1 +
P\Delta _2 + C\Delta _3 \} \nonumber\\
W(B^0_s\to \pi ^+K^-)  &=& -(T+P)(1+f_{27})\nonumber\\
&&-\frac{1}{5}\{ T(\Delta _2 -2 \Delta _1) + P \Delta _2 + 
C(3 \Delta _1-\Delta _2)\}\nonumber \\
W(B^0_s \to \pi ^0 \bar{K}^0 )&=& -\frac{1}{\sqrt{2}}(C-P)(1+f_{27})\nonumber
\\
&&+\frac{1}{5\sqrt{2}}\{ T(\Delta _2-2 \Delta _1) + P \Delta _2 +
C (3 \Delta _1 - \Delta _2) \} \nonumber \\
W(B^0_d \to \pi ^+ \pi ^-) & = &-(T+P)(1+f_{27})\nonumber \\
&&-\frac{1}{15}\{ T( -5\Delta _1 + 2\Delta _2 +\Delta _3 +\Delta _4  )
\nonumber \\ &&+P( \Delta _2 + \Delta _5 )
+C(  6\Delta _1 - 2 \Delta _2  - 3 \Delta _4 + \Delta _5)
\}\nonumber \\
W(B^0_d \to K^+K^- )&= &-\frac{1}{15}\{ T(-\Delta _1 + \Delta _2
-\Delta _3 -\Delta _4)\nonumber\\
&&+P(2\Delta _2-\Delta _5) + C(3\Delta _1 - \Delta _2  +
3 \Delta _4-\Delta _5) \} \nonumber \\
W(B^0_d \to \pi ^0 \pi ^0) &=&-\frac{1}{\sqrt{2}}(C-P)(1+f_{27})\nonumber \\
&&+\frac{1}{15\sqrt{2}}\{ T(-5\Delta _1 + 2\Delta _2 +\Delta _3 +
\Delta _4 )\nonumber \\
&&+P(\Delta _2 +\Delta _5) + C(6 \Delta _1 - 2\Delta _2 -3 \Delta _4
+\Delta _5)\}\nonumber \\
W(B^0_d \to K^0\bar{K}^0) &=&-P(1+f_{27})\nonumber \\
&&-\frac{1}{15}\{ T(4 \Delta _1-\Delta _2 - 2 \Delta _3 +\Delta _4)
\nonumber \\
\label{strangecons}
&&+P(\Delta _2 +\Delta _5 ) +C (3 \Delta _1 - \Delta _2  +
3 \Delta _4- \Delta _5)
\}
\end{eqnarray}
where the influence of FSI in the singlet channel is parametrized through
\begin{eqnarray}
\Delta _4&=&\frac{15}{4}(f_{8,8}-f_{27})\nonumber \\
\Delta _5&=&10(f_{8,8}-f_{27})+5f_{1,1}.
\end{eqnarray}

Similarly, in the $\Delta S =1$ sector (keeping only the dominant
$P'$, $T'$ in the FSI
contribution) we have:
\begin{eqnarray}
W(B^+ \to \pi ^+K^0) &=& -P'(1+f_{27})
\nonumber\\
&&-\frac{1}{5}\{ P'\Delta _2 + T'\Delta _1 \} \nonumber \\
W(B^+ \to \pi ^0 K^+) &=&\frac{1}{\sqrt{2}}(T'+C'+P')(1+f_{27})
\nonumber \\
&&+\frac{1}{5\sqrt{2}}\{ P' \Delta _2 +T'\Delta _1 \} \nonumber \\
W(B^0_d \to \pi ^- K^+ )&=& (T'+P')(1+f_{27}) \nonumber\\
&&+\frac{1}{5}\{ P' \Delta _2 +T' (\Delta _2 - 2 \Delta _1)
\} \nonumber \\
W(B^0_d \to \pi ^0 K^0 ) &=& \frac{1}{\sqrt{2}}(C'-P')(1+f_{27})\nonumber \\
&&-\frac{1}{5\sqrt{2}}\{ P' \Delta _2 +T' (\Delta _2 - 2 \Delta _1) \}
\nonumber \\
W(B^0_s \to \pi ^+ \pi ^-) &=& \frac{1}{15}\{ P'(2\Delta _2-\Delta _5)
+T'(-\Delta _1 +\Delta _2 - \Delta _3 -\Delta _4) \}\nonumber \\
W(B^0_s \to \pi ^0 \pi ^0) &=&-\frac{1}{15\sqrt{2}}
\{ P'(2\Delta _2 -\Delta _5) +T'(-\Delta _1 + \Delta _2 - \Delta _3
-\Delta _4)\} \nonumber \\
W(B^0_s \to K^+K^- ) &=&(T'+P')(1+f_{27})\nonumber\\
&&+\frac{1}{15}\{  P'(\Delta _2 +\Delta _5)+T'(-5\Delta _1 +2 \Delta _2
+\Delta _3 +\Delta _4) \}\nonumber \\
W(B^0_s \to K^0\bar{K}^0)&=&-P'(1+f_{27})\nonumber \\
&&-\frac{1}{15}\{  P'(\Delta _2 +\Delta _5)+ 
T'(4\Delta _1 -\Delta _2 - 2 \Delta _3 +\Delta _4) \}.\nonumber\\
\label{strangeviol}&&
\end{eqnarray}
Equations (\ref{strangecons},\ref{strangeviol}) quantify explicitly what is
already well known, i.e. that the
presence of significant FSI can be seen most directly 
in $B^0_d \to K^+K^-$ and $B^0_s \to \pi ^+ \pi ^-, \pi ^0 \pi ^0$.

For any FSI the above formulas satisfy the following three triangle
relations \cite{GHLR94}:
\begin{eqnarray}
W(B^+ \to \pi ^+ \pi ^0) &= &
\frac{1}{\sqrt{2}}W(B^0_s \to \pi^+ K^-)+W(B^0_s \to \pi ^0 \bar{K}^0)
\nonumber \\
W(B^0_s \to \pi^+ K^-)  &= & W(B^0_d \to \pi ^+ \pi ^-) + 
W(B^0 _d \to K^+K^-) \nonumber \\
\label{trianglecons}
W(B^0_d \to \pi ^0 \pi^0 &=& \frac{1}{\sqrt{2}}W(B^0_d \to K^+ K^-)
+ W(B^0_s \to \pi ^0 \bar{K}^0).
\end{eqnarray}
Alternatively, one of the three relations above may be replaced by the
isospin relation
\begin{eqnarray}
W(B^+ \to \pi ^+ \pi ^0) &=&
\frac{1}{\sqrt{2}}W(B^0_d \to \pi ^+ \pi ^-) + 
W(B^0_d \to \pi ^0 \pi^0)
\end{eqnarray}
(not independent of the previous three).

In the $\Delta S =1$ sector we have the following relations
\begin{eqnarray}
W(B^0_d \to \pi ^- K^+ ) + \sqrt{2}W(B^0_d \to \pi ^0 K^0 )  &
=& (T'+C')(1+f_{27})\nonumber \\
W(B^+ \to \pi ^+K^0) + \sqrt{2}W(B^+ \to \pi ^0 K^+)&
=& (T'+C')(1+f_{27}) \nonumber \\
W(B^0_s \to \pi ^+ \pi ^-) + \sqrt{2} W(B^0_s \to \pi ^0 \pi ^0)&
=& 0 \nonumber \\
\label{triangleviol}
W(B^0_s \to K^+K^- )+W(B^0_s \to \pi ^+ \pi ^-)&
=&W(B^0_d \to \pi ^- K^+ )
\end{eqnarray}
as discussed in \cite{GHLR94},
with the first two relations leading to
\begin{eqnarray}
&&W(B^0_d \to \pi ^- K^+ ) + \sqrt{2}W(B^0_d \to \pi ^0 K^0 )\nonumber \\
&=& W(B^+ \to \pi ^+K^0) + \sqrt{2}W(B^+ \to \pi ^0 K^+).
\label{quadrangle}
\end{eqnarray}
All these relations are FSI-independent.

Consequently, although the same five unknown complex parameters 
$\Delta _i$ (i=1,...5) enter
into both $\Delta S =0$ and $\Delta S =1$ sectors,
the number of all independent and in principle measurable data 
(i.e. decay widths) is not sufficient to determine all these parameters,
unless some additional input (like knowledge of sizes and relative
phases of $T,P,..$ and $T',P',...$ and/or $\Delta $'s,
assumption of higher-symmetry relations between $\Delta $'s, or justified
neglect of some terms) is accepted.

\section{Compatibility of quark-level parametrization with isospin}
In ref.\cite{GW97} it was argued that quark-diagram parametrization
in which $T'$ and $P'$ are given strong phases $\delta _{T'}$ and
$\delta _{P'}$ is not compatible with isospin invariance, unless
$\delta _{T'}-\delta _{P'}=\delta _{I=3/2}-\delta _{I=1/2} =0$
(see also ref.\cite{Wolf95}).

From the previous section we have
\begin{eqnarray}
W(B^+ \to \pi ^+K^0) &=& -P'(1+f_{27}+
\frac{1}{5} \Delta _2)  -\frac{1}{5}T'\Delta _1  \nonumber\\
\label{our}
W(B^0_d \to \pi ^- K^+ )&=& (T'+P')(1+f_{27}+\frac{1}{5}\Delta _2)
-\frac{2}{5} T' \Delta _1 \\
W(B^+ \to \pi ^0 K^+) &=&
\frac{1}{\sqrt{2}}(T'+P')(1+f_{27}+\frac{1}{5}\Delta _2)
+\frac{1}{5\sqrt{2}}T'(\Delta _1-\Delta _2)\nonumber \\
W(B^0_d \to \pi ^0 K^0)&=&
-\frac{1}{\sqrt{2}}P'(1+f_{27}+\frac{1}{5}\Delta _2)-
\frac{1}{5\sqrt{2}}T'(\Delta _2 - 2 \Delta _1).\nonumber
\end{eqnarray} 

This should be compared with the approach of \cite{GW97} which,
after adjustment to our notation, inclusion of weak phase $\gamma $ into
the definition of $T'$, $C'$, and the neglect of $C'$ terms, yields
(the first two equations below are Eqs.(6a,6b) of \cite{GW97},
$\delta = \delta_{3/2}-\delta_{1/2}$):
\begin{eqnarray}
A(B^+ \to \pi ^+K^0) &=& -  P' - \frac{1}{3}(1-e^{i\delta}) T'  
\nonumber\\
\label{Weyers}
A(B^0_d \to \pi ^- K^+ )&=& ( T'+P') +\frac{2}{3} (1 - e^{i\delta })
T' \\
A(B^+ \to \pi ^0 K^+)&=&\frac{1}{\sqrt{2}}(T'+P')-
\frac{1}{\sqrt{2}}\frac{2}{3}(1-e^{i\delta })T'\nonumber \\
A(B^0_d \to \pi ^0 K^0) &=& -\frac{1}{\sqrt{2}}P'-
\frac{1}{\sqrt{2}}\frac{2}{3}(1-e^{i\delta })T'.\nonumber
\end{eqnarray}
We see that the two sets of equations (\ref{our}) and 
(\ref{Weyers})
are identical if we make the following replacements
\begin{eqnarray}
P'(1+f_{27}+\frac{1}{5}\Delta _2 ) &\to & P' \nonumber \\
T'(1+f_{27}+\frac{1}{5}\Delta _2 ) &\to & T'
\end{eqnarray}
and  appropriately
choose $\Delta _1$  and $\Delta _2$, 
separately in each of the rightmost (and proportional to $T'$) terms in
Eqs.(\ref{our}).
The need for separate choices results from the oversimplified 
prescription for FSI used in \cite{GW97}: 
 a naive multiplication of quark-diagram amplitudes for
$I=1/2$ and $I=3/2$ by two different phases only. 
The latter prescription does
not allow for differences in various $I=1/2$ phases (e.g. from ${\bf 27}$
and from ${\bf 8}$, see also ref.\cite{ZWNG}), or
possible different changes in the absolute size of the
amplitudes.

Still, the general conclusion of \cite{GW97} is correct:
quark-diagram parametrization 
in which $P'$ and $T'$ are given different strong phases
is compatible with isospin symmetry in the $B \to \pi K$ decay channel
only if terms proportional to $T'\Delta _i$ (corresponding to
$(1-e^{i\delta })T'$) 
are neglected.
As one expects that $|T'| < |P'|$,
neglecting $T'\Delta _i$ terms might seem a reasonable approximation for 
strangeness-violating $B \to \pi K$ decays.
However, when $\Delta S =1$ decays $B \to \pi \pi, K\bar{K}$ are also
considered, a glance at Eqs(\ref{strangeviol}) shows that
different modifications of $P'$ are needed there.

For $\Delta S =0$ decays, the dominant FSI-induced correction
terms should be proportional to $T$. Eqs(\ref{strangecons})
show then that FSI-induced terms proportional to $T$ enter different
amplitudes with different
coefficients and no universal renormalization of quark-level
amplitudes $T$, $P$, $C$ can work.
In general, therefore, 
parametrization of FSI effects by endowing quark-diagram
amplitudes $T$, $P$, $C$
with additional universal phases cannot take 
the whole complexity of FSI into account.

\section{Restriction to leading FSI corrections}
If final-state interactions may be treated as a correction to the
direct SD amplitudes, it seems natural to keep  leading terms only
in such a correction. 
Assuming then that $\Delta _i$ are all of similar sizes,
we may neglect in Eqs.(\ref{strangecons},\ref{strangeviol}) all
FSI-induced terms but the leading ones, i.e. those
proportional to $T$ and $P'$. Thus,
the $\Delta S =0$ decay amplitudes depend  on four
$\Delta _i$ ($\Delta _5$ drops out), while the $\Delta S =1$
amplitudes on two $\Delta _i$: $\Delta _2$ and $\Delta _5$.

In the $\Delta S = 0$ sector, 
with amplitudes still depending on four $\Delta _i$, 
no relations between amplitudes in addition to those of 
Eq.(\ref{trianglecons})
are generated.
The number of undetermined parameters is too large to permit
their clear-cut determination from data.
Thus, additional input is necessary.

In the $\Delta S =1$ sector
it is instructive to rewrite the amplitudes
 in terms of redefined
quark-diagram amplitudes:
\begin{eqnarray}
\tilde{T'}&\equiv &T'(1+f_{27})\nonumber\\
\tilde{C'}&\equiv &C'(1+f_{27})\nonumber\\
\label{primed}
\tilde{P'}&\equiv &P'(1+f_{27}+\frac{1}{5}\Delta _2)
\end{eqnarray}
and
\begin{equation}
\tilde{\Delta } \equiv \frac{1}{15}(2\Delta _2-\Delta _5)/(1+f_{27}+
\frac{1}{5}\Delta _2).
\end{equation}
One then obtains
\begin{eqnarray}
W(B^+ \to \pi ^+ K^0)&=&-\tilde{P'}\nonumber\\
W(B^+ \to \pi ^0 K^+)&=&\frac{1}{\sqrt{2}}(\tilde{T'}+\tilde{C'}
+\tilde{P'})\nonumber\\
W(B^0_d \to \pi ^- K^+) &=& \tilde{T'}+\tilde{P'}\nonumber\\
W(B^0_d \to \pi ^0 K^0) &=&\frac{1}{\sqrt{2}}(\tilde{C'}-\tilde{P'})
\nonumber\\
W(B^0_s \to \pi ^+ \pi ^-) &= &\tilde{P'} \tilde{\Delta}\nonumber \\
W(B^0_s \to \pi ^0 \pi ^0) &= &-\frac{1}{\sqrt{2}}
\tilde{P'} \tilde{\Delta}\nonumber \\
W(B^0_s \to K^+K^-) &=&\tilde{T'}+\tilde{P'}-\tilde{P'}\tilde{\Delta }
\nonumber\\
W(B^0_s \to K^0\bar{K}^0) &=&-\tilde{P'}+\tilde{P'}\tilde{\Delta }.
\end{eqnarray}
Note that 
the first four equations above have the structure used in
the SD quark-diagram approach:
the FSI effects can be identified only with additional
 help from $B^0_s$ decays.
With eight decays and four parameters ($\tilde{T'}$,
$\tilde{P'}$, $\tilde{C'}$, $\tilde{\Delta}$) there are four relations
between the amplitudes. In addition to the three relations of 
Eqs.(\ref{triangleviol},\ref{quadrangle}),
we have one new relation involving $B^0_s \to K^0 \bar{K}^0$:
\begin{equation}
W(B^+ \to \pi ^+ K^0)+W(B^0_s \to \pi ^+ \pi ^-)=
W(B^0_s \to K^o \bar{K}^0).
\end{equation}
This relation yields information on the phase of the FSI-related
parameter $\tilde{\Delta }$.
Note that the ratio 
$|\sqrt{2}W(B^0_s \to \pi ^+ \pi ^-)/W(B^+ \to \pi ^+K^0)|$
measures the (relative) size of observable FSI effects.

\section{Relating $\Delta S =0$ and $\Delta S=1$ amplitudes}
In the
SD quark-diagram approach
the $\Delta S =0$ and $\Delta S =1$ decay amplitudes are related.
Consequently, simultaneous analyses of these 
amplitudes have been considered as a means to
provide important tests of the approximations made in the SD approach,
and as a way to extract weak angle $\gamma $.
An important question is how such analyses are affected by FSI effects.

It appears that
 rescattering may  upset 
 expectations related to $s \leftrightarrow d$
 flavor  $U$-spin reflection arguments
  \cite{GronauUspin}.
Consider for example
the amplitudes for the four decays 
\begin{eqnarray}
B^+   & \to & K^+ \bar{K}^0 \nonumber\\
B^0_s & \to & \pi ^+ K^- \nonumber\\
B^+   & \to & \pi ^+ K^0    \nonumber\\
\label{PTandDelta}
B^0_d & \to & \pi ^- K^+. 
\end{eqnarray} 
Introducing 
\begin{eqnarray}
\tilde{P}&=& P(1+f_{27}+\frac{1}{5}\Delta _2)\nonumber\\
\tilde{T}&=& T(1+f_{27}+\frac{1}{5}\Delta _2)\nonumber\\
\tilde{\Delta _1} &=& \Delta _1/(1+f_{27}+\frac{1}{5}\Delta _2)
\nonumber \\
\label{eq28}
\tilde{\Delta _2} &=& \Delta _2/(1+f_{27}+\frac{1}{5}\Delta _2)
\end{eqnarray}
in addition to $\tilde{P}'=P'(1+f_{27}+\Delta _2/5)$ (Eq.(\ref{primed})),
the amplitudes
for the first two ($\Delta S =0$) decays in Eq.(\ref{PTandDelta}) 
may be
reexpressed 
as
\begin{eqnarray}
W(B^+ \to K^+ \bar{K}^0)  &=& - \tilde{P} 
- \frac{1}{5}\tilde{T}\tilde{\Delta _1} 
\nonumber\\
\label{tildeunprimed}
W(B^0_s \to \pi ^+ K^-)   &=& 
-\tilde{P}-\tilde{T}
+\frac{2}{5}
\tilde{T}\tilde{\Delta _1 }
\end{eqnarray}
when the FSI-induced terms proportional to $C$ are neglected.
Note that we have kept terms of order $P\Delta _2$
even though they represent nonleading FSI effects.
Similarly, we could have kept nonleading terms of order $T'\Delta _2$
in the definition of $T'$ in Eq.(\ref{primed}), i.e.
\begin{equation}
\label{Ttildeprimed}
\tilde{T}' = T'(1+f_{27}+\frac{1}{5}\Delta _2)
\end{equation}
so that
\begin{equation}
\label{Ttildeandprimed}
\tilde{T}=\frac{1}{\lambda}\tilde{T}'
\end{equation}
where $\lambda \approx 0.22 $ is the parameter 
in the Wolfenstein's parametrization
of the CKM matrix.

From Eqs.(\ref{strangeviol})
the two $\Delta S=1$ decays of Eqs.(\ref{PTandDelta})
are then described by
\begin{eqnarray}
W(B^+ \to \pi ^+ K^0 )    &=& -\tilde{P}'-\frac{1}{5}\tilde{T}'
\tilde{\Delta _1}
\nonumber \\
W(B^0_d \to \pi ^- K^+)   &=& \tilde{P}'+\tilde{T}'-\frac{2}{5}\tilde{T}'
\tilde{\Delta _1}
\label{tildeprimed}
\end{eqnarray}

Corrections from electroweak penguin diagrams to the right-hand sides
of Eqs.(\ref{tildeunprimed}) (Eqs.(\ref{tildeprimed})) are proportional to
$P^c_{EW}/3$ and $-2P^c_{EW}/3$  ($P'^c_{EW}/3$ and  $-2P'^c_{EW}/3$)
respectively. 
(Actually, using the substitutions
$T \to T+P^c_{EW}$, $P \to P-P^c_{EW}/3$, $C \to C+P_{EW}$,  and the
analogous ones for the $\Delta S =1$ transitions,
we could have started our calculations from SD
amplitudes corrected for electroweak penguins.)
Since one expects that
$|P^c_{EW}|<|E|,|A|,|P_{EW}|<|C|,|P|<|T|$, and
$|P'^c_{EW}|< 0.05|P'|<|T'|\approx (0.1~ {\rm to}~ 0.2)|P'|$ 
\cite{elpenguins} (see also \cite{EWpenguins}), 
any such contributions
have to be neglected in our approximation.
Only the $P'_{EW}$ terms (of order $T'$) should be included in
the non-FSI-induced terms in Eqs.(\ref{PTandDelta}).
However, in FSI-induced terms the corrections from the $P'_{EW}$
should be neglected if those from $T'$ are.
Thus, when terms of order $\tilde{T} \tilde{\Delta _1} =\tilde{T}'
\tilde{\Delta _1}
/\lambda $  (in $B^+ \to K^+ \bar{K}^0$ and $B^0_s \to \pi ^+K^-$)
are kept, 
but those of order $\tilde{T}' \tilde{\Delta _1} $ 
(in $B^+ \to \pi ^+ K^0$ and $B^0_d \to \pi ^- K^+$) are neglected,
 our final form of Eqs.(\ref{tildeunprimed},\ref{tildeprimed}) is
\begin{eqnarray}
W(B^+ \to K^+ \bar{K}^0)  =& 
 -\tilde{P}-\frac{1}{5}\tilde{T}\tilde{\Delta }_1 
&=  - \tilde{P} 
- \frac{1}{5}\frac{1}{\lambda }\tilde{T}' \tilde{\Delta _1} 
\nonumber\\
W(B^0_s \to \pi ^+ K^-)  = &  -\tilde{P}-\tilde{T}+\frac{2}{5}\tilde{T}
\tilde{\Delta _1}
& = 
- \tilde{P}-\frac{1}{\lambda }\tilde{T}'
+\frac{2}{5}\frac{1}{\lambda }
\tilde{T}'\tilde{\Delta _1 }
\nonumber\\
W(B^+ \to \pi ^+ K^0 )   = &  -\tilde{P}' &
\nonumber \\
W(B^0_d \to \pi ^- K^+)  =& \tilde{P}'+\tilde{T}'.&
\label{finalFalkGR}
\end{eqnarray}
with $\tilde{P}$, $\tilde{T}$ defined in Eq.(\ref{eq28}),
$\tilde{P}'$ defined in Eq.(\ref{primed})
and $\tilde{T}'$ defined in Eq.(\ref{Ttildeprimed}).
In Eqs.(\ref{finalFalkGR}) a part of rescattering effects is included
into the definition of effective "penguin" and "tree" amplitudes
$\tilde{P}$,
$\tilde{P}'$ and 
$\tilde{T}$, $\tilde{T}'$ through a common multiplicative
factor of $(1+f_{27}+\Delta _2/5)$.
It is only
the term
$-\frac{1}{5}
\tilde{T}\tilde{\Delta _1}$ in the expression for
$W(B^+ \to K ^+ \bar{K}^0)$ 
(and a similar one in $W(B^0_s \to \pi ^+ K^-)$)
which represents "visible" FSI effects
 (i.e. those not removable through a redefinition of $P$, $T$ amplitudes).
 This term may  influence 
the equality
\begin{equation}
\label{annihilation}
W(B^+ \to K^+ \bar{K}^0)=-\lambda W(B^+ \to \pi ^+ K^0)
\end{equation}
obtained (for SU(3) symmetric $P$ and $P'$) either
when charming penguins are dominant, or
in SD approaches when $\beta \approx 0$ (see later).
Comparison of $B^+ \to K^+ \bar{K}^0$
and $B^+ \to \pi ^+ K^0$ 
was considered as a test for the presence
of the contribution from the annihilation diagram 
or FSI effects \cite{Rosnerrev,Falk98}. 
Indeed, the relative size of the FSI-generated correction
term to $\tilde{P}$ in $W(B^+ \to K^+ \bar{K}^0) $ is proportional
to $\frac{1}{5}|T/P|$
and, with $|P/T|\approx 0.3\pm 0.1$, it might be sizable.
Note that by including two terms of different weak phases,
the first of Eqs.(\ref{finalFalkGR}) explicitly indicates
the appearance of a rescattering-induced CP-violating asymmetry
$\Gamma (B^+ \to K^+\bar{K}^0)-\Gamma (B^- \to K^- K^0)$.
Great
importance of $W(B^+ \to K^+ \bar{K}^0)$ for gathering information on
rescattering effects was also noted  in \cite{FSIK+K-}.
The present approach places such considerations in a 
framework which quantifies the connections between all FSI effects in
 $B$ decays into 
$\pi \pi,~ K\bar{K}$, and $ \pi K$.

Qualitatively, violation of equality (\ref{annihilation})
by FSI effects
may be understood as follows.
The amplitude $W(B^+\to K^+\bar{K}^0)$ receives contributions
from inelastic intermediate states with flavour content
$"PV" = "\pi ^+ \omega _8"$, $"PV" = "\pi ^+ \omega _1"$, 
$"PV" = "\eta _8 \rho ^+"$, etc.
These amplitudes involve tree amplitudes  proportional 
to the SD tree amplitude $T$ 
(in addition to the amplitudes proportional to
$P$, etc. ). The approximations
involved when deriving the first of Eqs.
(\ref{tildeunprimed}) leave the $T \Delta$ term  as the only sizable
FSI-induced term (as $|T| > |P| \approx |C| > ...$).
On the other hand, although the FSI-induced 
corrections to $W(B^+ \to \pi ^+ K^0)$
 also  contain (compare Eqs.(\ref{tildeprimed})) analogous terms
proportional to the SD tree amplitude $T'$
(originating from inelastic intermediate 
states "$PV"="\pi ^0 K^{*+}$", "$PV"="K^+\omega _8$" etc.), 
the approximations
involved neglect these terms on account of $|P'|>|T'|>|C'|...$.

Combined analysis of decays $B^+ \to \pi ^+ K^0$, $B^0_d \to \pi ^- K^+$,
and $B^0_s \to \pi ^+ K^-$ 
(together with their CP
counterparts) was proposed in ref.\cite{GR482} as a means to provide
information on the value of the $CP$-violating angle $\gamma $.
From the form of the expressions for relevant amplitudes in the presence
of FSI (the last three equations in Eqs.(\ref{finalFalkGR})),
we see that rescattering might affect 
the determination of $\gamma $ (see also \cite{GP}):
the FSI-induced term in the $B^0_s \to \pi^- K^+$ amplitude
 is of the order of
 $\frac{2}{5}|T/P|$ of the penguin
 amplitude $P$, i.e. twice the size
 of a similar term in $B^+ \to K^+\bar{K}^0$.
 Thus, if rescattering effects in $B^+ \to K^+\bar{K}^0$ 
 are substantial, one should seriously
 worry about the FSI corrections to the method of ref.\cite{GR482}.
 
 In ref.\cite{GR482}, using unitarity of the CKM matrix, i.e.
$V^*_{tb}V_{ti}=-V^*_{cb}V_{ci}-V^*_{ub}V_{ui}$~, only the
$-V^*_{cb}V_{ci}$ part of the penguins is included into
the redefined penguins $p$ and $p'$:
\begin{eqnarray}
P=&p
 \left( 1-\frac{\sin \beta}{\sin (\beta +\gamma )}e^{i\gamma } \right)
 &=p\frac{\sin \gamma}{\sin (\beta +\gamma)}e^{-i\beta}\nonumber \\
 \label{Pp}
 P'=&p' \left( 1+ \lambda ^2 \frac{\sin \beta}{\sin (\beta + \gamma)}
 e^{i\gamma}\right)&
\end{eqnarray}
with
\begin{equation}
p=-\lambda p'
\end{equation}
 while the $-V^*_{ub}V_{ui}$ parts 
are absorbed into the redefined tree amplitudes $t$, $t'$
\begin{eqnarray}
T&=&t+p\frac{\sin \beta}{\sin (\beta + \gamma )}e^{i \gamma }\nonumber \\
\label{Tt}
T'&=&t'-\lambda ^2 p'\frac{\sin \beta}{\sin (\beta +\gamma)}
e^{i\gamma}
\end{eqnarray}
with
\begin{equation}
\label{}
t=\frac{1}{\lambda }t'
\end{equation}
 The approximation of ref.\cite{GR482} consists in neglecting the 
 $\lambda ^2$ terms in the expression relating $P'$ and 
 $p'$, i.e. it corresponds to
 $\beta \to 0$ \cite{CW}.
 
 With FSI taken into account, by
 replacing the $\tilde{T}'\tilde{\Delta _1}$ terms with
 $\tilde{t}'\tilde{\delta _1}\equiv 
 \tilde{T}'\tilde{\Delta _1}$,
 where $\tilde{t}'$ is related to $\tilde{T}'$
 through an analogon of Eqs(\ref{Tt}),
 and with $\tilde{p}'=p'(1+f_{27}+\Delta _2/5)$,
 we have
 \begin{eqnarray}
 W(B^+ \to K^+\bar{K}^0) &= &\lambda \tilde{p}'
 \left( 1-\frac{\sin \beta}{\sin (\beta + 
 \gamma)}e^{i\gamma}\right)
 -\frac{1}{5}\frac{1}{\lambda}\tilde{t}'
 \tilde{\delta _1} \nonumber \\
 W(B^0_s \to \pi ^+ K^-) &= &\lambda \tilde{p}'
 -\frac{1}{\lambda }\tilde{t}'+
 \frac{2}{5}\frac{1}{\lambda}\tilde{t}'
 \tilde{\delta _1} \nonumber \\
 W(B^+ \to \pi ^+ K^0)&=& -\tilde{p}'
 \left( 1+ \lambda ^2 \frac{\sin \beta}{\sin (\beta +\gamma )} 
 e^{i\gamma }\right) \nonumber \\
 \label{final}
 W(B^0_d \to \pi ^- K^+ ) &=& \tilde{p}'+\tilde{t}'
 \end{eqnarray}
 If the charming penguins of ref.\cite{Martinelli} are substantial, 
 they may
 be included into the definition of $\beta $-independent 
 parts of redefined penguins
 above, effectively suppressing
 the $\beta $-dependent
 parts (and leading to
 Eq(\ref{annihilation})).
 
 In the following formulas we accept that $\beta $ is small, so that
 terms proportional to $\sin \beta$ may be neglected; 
 in reality, a nonzero value of
 $\beta $ would have to be used in any attempt to extract the
 angle $\gamma $ from data on the basis of Eqs(\ref{final})
 \cite{CW}.
 
  The equality $A_0=-A_s$, expected to hold 
  (for any $\beta $) in SU(3) \cite{GR482}
  between the CP-violating rate pseudo-asymmetries
 \begin{equation}
 \label{A0def}
 A_0 \equiv \frac{\Gamma (B^0_d \to K^+ \pi ^-)-
 \Gamma (\bar{B}^0_d \to K^-\pi ^+)}
 {\Gamma (B^+ \to K^0 \pi ^+) + \Gamma (B^- \to \bar{K}^0 \pi ^-)}
 \end{equation}
 and
 \begin{equation}
 \label{Asdef}
 A_s \equiv \frac{\Gamma (B^0_s \to K^- \pi ^+)-
 \Gamma (\bar{B}^0_s\to K^+\pi ^-)}
 {\Gamma (B^+ \to K^0 \pi ^+) + \Gamma (B^- \to \bar{K}^0 \pi ^-)},
 \end{equation}
 may be affected by FSI
 even when the latter is SU(3) symmetric.
 Indeed, using
Eqs.(\ref{final}) one derives (for $\beta \approx 0$)
\begin{equation}
\label{A0}
A_0=-2r \sin \delta \sin \gamma
\end{equation}
and
\begin{equation}
\label{As}
A_s=2r\kappa \sin (\delta +\epsilon) \sin \gamma
\end{equation}
where $\delta $ ($\gamma $) denotes relative strong (weak) 
phase of $t'$($\tilde{t}'$) 
with respect to $p'$ ($\tilde{p}'$), 
$r=|\tilde{t}'/\tilde{p}'|=|t'/p'|$,
and (with $\kappa = |\kappa |$)
\begin{equation}
\label{kappaepsilon}
1-\frac{2}{5}\tilde{\delta }_1 \equiv \kappa e ^{i\epsilon}.
\end{equation}
Since from Eqs.(\ref{final}) the charge-averaged ratios 
\begin{equation}
 \label{Rdef}
 R \equiv \frac{\Gamma (B^0_d \to K^+ \pi ^-)+
 \Gamma (\bar{B}^0_d \to K^-\pi ^+)}
 {\Gamma (B^+ \to K^0 \pi ^+) + \Gamma (B^- \to \bar{K}^0 \pi ^-)}
 \end{equation}
 and
 \begin{equation}
 \label{Rsdef}
 R_s \equiv \frac{\Gamma (B^0_s \to K^- \pi ^+)+
 \Gamma (\bar{B}^0_s\to K^+\pi ^-)}
 {\Gamma (B^+ \to K^0 \pi ^+) + \Gamma (B^- \to \bar{K}^0 \pi ^-)}
 \end{equation}
 are given
by
\begin{eqnarray}
\label{R}
R   &=& 1+r^2+2r \cos \delta \cos \gamma  \\
\label{Rs}
R_s &=& \lambda ^2 +\frac{r^2}{\lambda ^2}-2r \kappa \cos \gamma \cos
(\delta + \epsilon ),
\end{eqnarray}
there are now four equations (\ref{A0}),(\ref{As}),(\ref{R}),(\ref{Rs}) 
for five unknowns ($r$, $\gamma$, $\delta $, $\kappa $, $\epsilon $).
If $\epsilon \ll \delta$, the four equations may be solved after neglecting
$\epsilon $. For $\epsilon $ of order $\delta $,
additional constraints would be needed.
The ratios $(\Gamma (B^+ \to K^+\bar{K}^0)
\pm \Gamma (B^- \to K^-K^0))/
(\Gamma (B^+ \to K^0 \pi ^+) + \Gamma (B^- \to \bar{K}^0 \pi ^-))$ 
may be expressed in terms of $r$, .., $\epsilon$ (and $\beta $
when its nonzero value is taken into account),
and seem to provide 
such constraints.  Thus, if $A_0\ne -A_s$, their
usefulness would have to be studied.
Such an analysis requires a detailed consideration of SU(3) breaking
which is outside the scope of this paper.

 Similar effects of apparent SU(3) breaking
 are observed for other pairs of U-spin-related decays. 
 According to Eqs.(\ref{strangecons},\ref{strangeviol}),
 when $E$ and $PA$ ($E'$ and $PA'$) SD amplitudes are neglected, the 
 processes $B^0_d \to K^+K^-$ and
 $B^0_s \to \pi ^+ \pi ^-$, related to one another
 by this reflection, are described by rescattering-induced
 amplitudes:
 \begin{eqnarray}
 W(B^0_d \to K^+K^-)&=& 
 -\frac{1}{15}\{
 T(-\Delta _1+\Delta _2-\Delta _3 -\Delta _4)
 +P(2\Delta _2-\Delta _5)+...
 \}
 \nonumber\\
 W(B^0_s \to \pi^+ \pi^- )&=&
 \frac{1}{15}\{
 T'(-\Delta _1+\Delta _2-\Delta _3 -\Delta _4)
 +P'(2\Delta _2-\Delta _5)
 \}
 \end{eqnarray}
 If $P/T$ were equal to $P'/T'$, we would indeed expect 
 for $|T/T'|=|V_{ud}/V_{us}|$ that
 \begin{equation}
 \label{UspinKKpipi}
 \frac{\Gamma (B^0_d \to K^+K^-)}{\Gamma (B^0_s \to \pi^+ \pi^-)}
 =\left| \frac{V_{ud}}{V_{us}}\right| ^2,
 \end{equation}
 as obtained in SD approaches.
 However, as one expects  
 that $|T'/T|\approx |P/P'|$
 with dominant $T$- and $P'$- terms, relation (\ref{UspinKKpipi})
 may be violated. Thus, 
 Eq.(\ref{UspinKKpipi}) may help distinguish
 between rescattering effects and
 genuine short-distance $E$ and $PA$
 contributions.
 
 A look at Eqs.(\ref{strangecons},\ref{strangeviol}) shows 
 that the method of ref.\cite{Fleischer}, based on the
 U-spin-related decays $B^0_d \to \pi ^+ \pi ^-$
and $B^0_s \to K^+ K^-$, is also affected by rescattering.
Indeed, keeping only the dominant FSI-induced terms 
(i.e. those proportional to $T$ and $P'$) introduces
 two unrelated linear combinations of 
$\Delta $'s into the game. Thus, FSI-induced modifications of this method
are less easily controlled than those of ref.\cite{GR482}.

 When
 specific models for rescattering relations 
 (and thus, definite relations between $\Delta $'s) are
considered, 
 further relations between FSI-induced corrections to various decays
 should appear.
 The analysis of such models and their predictions
 is outside the scope of this paper.

\section{Conclusions}
In this paper we analysed 
the influence of SU(3)-symmetric 
inelastic rescattering onto the predictions 
of  short-distance  quark-diagram approach to $B $ decays
into two pseudoscalars $PP$
when the tree and penguin amplitudes are assumed dominant.
Final-state interactions were described with
the help of a few parameters corresponding to all possible
SU(3)-symmetric forms of inelastic rescattering into $PP$.
We found that the combined set of experimental data on all
 $B \to \pi \pi ,~ K\bar{K},~\pi K$ decays is not sufficient
 to determine all relevant FSI-related parameters.
 Still, some important 
 information on inelastic FSI effects may be extracted
 from the data. 
 Apart from providing explicit
 expressions for the amplitudes of the FSI-driven decays $B^0_d \to K^+K^-$,
 $B^0_s \to \pi ^+ \pi ^-$, and $B^0_s \to \pi ^0 \pi ^0$,
 it was shown that the $\Delta S =1$ decays
 may provide quantitative information on the 
 magnitude and phase of the single
 FSI-indicating effective parameter appearing in this sector.
 FSI-induced modification of the
  connection between $B^+ \to K^+ \bar{K}^0$ and $B^+ \to \pi ^+ K^0$
 amplitudes was also given explicitly.
 Furthermore, it was shown that rescattering 
 affects the analyses of
 U-spin-related decays.
 In particular, by modifying
 the SD prescription for the
 amplitudes of $B^+ \to \pi ^+ K^0$, $B^0_d \to \pi ^- K^+$, and
 $B^0_s \to \pi ^+ K^-$ decays,  FSI may affect the 
  method of determining
 the CP-violating angle $\gamma $,
 which uses the corresponding decay rates as input.
 Deviation from equality $A_0=-A_s$ may indicate SU(3) breaking
 induced by SU(3)-symmetric FSI effects.

\section{Acknowledgements}
This work was supported in part by the
Polish State Committee for Scientific Research 
grant 5 P03B 050 21.

\section{Appendix}
\subsection{Mesons}

\begin{tabular}{lll}
\phantom{XXXXXXXXXXXXXXXXXXX}&\phantom{XXXXXXXXXXX}&\phantom{XXXXXXXXXX}\\
$\pi ^+ = -u\bar{d}$ & $K^+ =u\bar{s}$& $B^+ =u\bar{b} $ \\
$\pi ^0 =(u\bar{u}-d\bar{d})/\sqrt{2}$&$K^0=d\bar{s}$&$B^0_d=d\bar{b}$ \\
$\pi ^-=d\bar{u}$ &$K^-=s\bar{u}$& $B^0_s=s\bar{b}$ \\
$\eta _8=(u\bar{u}+d\bar{d}-2s\bar{s})/\sqrt{6}$&$\bar{K}^0=-s\bar{d}$&
\\
$\eta _1=(u\bar{u}+d\bar{d}+s\bar{s})/\sqrt{3}$&&\\
\end{tabular}

Analogous conventions hold for vector- and other mesons.
In the following we denote $K=(K^+ ,K^0)$, $\bar{K} =(\bar{K} ^0, K^-)$.

\subsection{Two-meson $PP$ states}

Two-meson $PP$ states of definite isospin $I$ are denoted as $(ab)_I$. 
Since the charge of state $(ab)_I$ must correspond to the charge of 
the decaying B-meson, the value of charge is suppressed whenever
this does not lead to ambiguity. 
States $(ab)_I$
with mesons $a$ and $b$ in definite charge states 
are defined according
to the following example for charge $Q=+1$:
\begin{equation}
\label{eq:ex}
(\pi \pi)_2=+\{ \pi ^+ \pi ^0\}
\end{equation}
where $\{a^{q_1}b^{q_2}\}$ denotes a properly symmetrized state, 
i.e. $\{a^{q_1}b^{q_2}\}= (a^{q_1}b^{q_2}+b^{q_2}a^{q_1})/\sqrt{2}$. 
If $b^{q_2}=a^{q_1}$, $\{a^qa^q\}=a^qa^q$. (All relations of type
(\ref{eq:ex}) have a positive sign on the right-hand side).
States in which mesons $a$ and $b$ are not in definite charge states are
represented as linear combinations of states with definite charges of mesons
$a$ and $b$.
All relevant states of given charge, strangeness 
and definite isospin are listed below.

\noindent
a) Strangeness $S=0$, charges $Q=+1,0$
\begin{eqnarray}
(\pi \pi )_2 & = & \left\{
\begin{array}{ll}
\{ \pi ^+ \pi ^0\} & \mbox{if $Q=+1$}\\
(\{\pi ^+\pi ^-\}+\sqrt{2}\{\pi ^0 \pi ^0\})/\sqrt{3}\phantom{xxx}&
\mbox{if $Q=0$}
\end{array}
\right.
\nonumber \\
(K \bar{K} )_1 & = & \left\{
\begin{array}{ll}
\{ K^+ \bar{K} ^0\} &\mbox{if $Q=+1$}\\
(\{K^+K^-\}+\{K^0\bar{K}^0\})/\sqrt{2} \phantom{xxx}
& \mbox{if $Q=0$}
\end{array}
\right.
\nonumber \\
\left.
\begin{array}{r}
(\pi \eta _8)_1 \\ (\pi \eta _1)_1
\end{array} 
\right\}  &&
\mbox{both charges} \nonumber \\
(\pi \pi)_0 &= &(\sqrt{2}\{\pi ^+ \pi ^-\}-\{\pi ^0 \pi ^0\})/\sqrt{3} 
\nonumber \\
(K\bar{K})_0 &= &(\{K^+K^-\}-\{K^0\bar{K}^0\})/\sqrt{2} \nonumber \\
(\eta _8 \eta _8)_0 && \nonumber \\
(\eta _1 \eta _1)_0 && \nonumber \\
(\eta _8 \eta _1)_0 &&
\end{eqnarray}

\noindent
b) Strangeness $S=+1$, charges $Q=+1,0$
\begin{eqnarray}
(\pi K)_{3/2}&=&\left\{
\begin{array}{ll}
\sqrt{\frac{2}{3}}\{\pi ^0 K^+\}+\frac{1}{\sqrt{3}}\{\pi ^+ K^0\}
& \mbox{if $Q=+1$} \\
\frac{1}{\sqrt{3}}\{\pi ^- K^+\}+
\sqrt{\frac{2}{3}}\{\pi ^0 K^0\} \phantom{xxx}
& \mbox{if $Q=0$}
\end{array}
\right. \nonumber \\
(\pi K)_{1/2} &=&\left\{
\begin{array}{ll}
\frac{1}{\sqrt{3}}\{\pi ^0 K^+\}-
\sqrt{\frac{2}{3}}\{\pi ^+ K^0\} & \mbox{if $Q=+1$} \\
\sqrt{\frac{2}{3}}\{\pi ^- K^+\}-\frac{1}{\sqrt{3}}\{\pi ^0 K^0\}
\phantom{xxx}
& \mbox{if $Q=0$}
\end{array}
\right.
\nonumber \\
\left.
\begin{array}{r}
(\eta _8 K)_{1/2} \\
(\eta _1 K)_{1/2}
\end{array}
\right\}&& \mbox{both charges}
\end{eqnarray}

\noindent
c) Strangeness $S=-1$, charge $Q=0$
\begin{eqnarray}
(\pi \bar{K})_{3/2} &=&\frac{1}{\sqrt{3}}\{\pi ^+ K^-\}+
\sqrt{\frac{2}{3}}\{\pi ^0 \bar{K}^0\}\nonumber \\ 
(\pi \bar{K})_{1/2}&=&
\sqrt{\frac{2}{3}}\{\pi ^+ K^-\}-\frac{1}{\sqrt{3}}\{\pi ^0 \bar{K}^0\}
\nonumber \\
(\eta _8 \bar{K})_{1/2} && \nonumber \\
(\eta _1 \bar{K})_{1/2}
&& 
\end{eqnarray}

\subsection{States in definite SU(3) representations}

Notation used: ({\bf SU(3) multiplet}, isospin)

\noindent
a) Strangeness $S=0$

Isospin $2$, charges $Q=+1,0$
\begin{eqnarray}
({\bf 27},2)&=&(\pi \pi )_2
\end{eqnarray}

Isospin $1$, charges $Q=+1,0$
\begin{eqnarray}
\left[
\begin{array}{r}
({\bf 27},1)\\({\bf 8},1)
\end{array}
\right]
&=&\frac{1}{\sqrt{5}}
\left[
\begin{array}{rr}
\sqrt{3} & -\sqrt{2}\\
\sqrt{2} & \sqrt{3}
\end{array}
\right]
\left[
\begin{array}{r}
(\pi \eta _8)_1 \\ (K \bar{K} )_1
\end{array}
\right]
\nonumber\\
({\bf 8}',1)\phantom{xi}&=&(\pi \eta _1)_1
\end{eqnarray}

Isospin $0$, charge $Q=0$
\begin{eqnarray}
\left[
\begin{array}{r}
({\bf 27},0)\\ ({\bf 8},0) \\ ({\bf 1},0)
\end{array}
\right]
&=&
\left[
\begin{array}{rrr}
\frac{1}{2\sqrt{10}}&\sqrt{\frac{3}{10}}&-\frac{3\sqrt{3}}{2\sqrt{10}}\\
\sqrt{\frac{3}{5}}&\frac{1}{\sqrt{5}}&\frac{1}{\sqrt{5}}\\
-\frac{\sqrt{3}}{2\sqrt{2}}&\frac{1}{\sqrt{2}}&\frac{1}{2\sqrt{2}}
\end{array}
\right]
\left[
\begin{array}{r}
(\pi \pi)_0 \\ (K \bar{K} )_0 \\ (\eta _8 \eta _8)_0
\end{array}
\right]
\nonumber \\
({\bf 1}',0)\phantom{xi}&=&(\eta _1 \eta _1 )_0
\end{eqnarray}

\noindent
b) Strangeness $S=+1$, charges $Q=+1,0$
\begin{eqnarray}
({\bf 27},3/2)\phantom{xi}&=&(\pi K)_{3/2}\nonumber\\
\left[
\begin{array}{r}
({\bf 27},1/2) \\ ({\bf 8},1/2)
\end{array}
\right]
&=&\frac{1}{\sqrt{10}}
\left[
\begin{array}{rr}
1 & 3\\
3 & -1
\end{array}
\right]
\left[
\begin{array}{r}
(\pi K)_{1/2} \\ (\eta _8 K)_{1/2}
\end{array}
\right]\nonumber \\
({\bf 8}',1/2)\phantom{xi}&=&(\eta _1 K)_{1/2}
\end{eqnarray}

\noindent
c) Strangeness $S=-1$, charge $Q=0$
\begin{eqnarray}
({\bf 27},3/2)\phantom{xi}&=&(\pi \bar{K})_{3/2}\nonumber\\
\left[
\begin{array}{r}
({\bf 27},1/2) \\ ({\bf 8},1/2)
\end{array}
\right]
&=&\frac{1}{\sqrt{10}}
\left[
\begin{array}{rr}
1 & 3\\
3 & -1
\end{array}
\right]
\left[
\begin{array}{r}
(\pi \bar{K})_{1/2} \\ (\eta _8 \bar{K})_{1/2}
\end{array}
\right]\nonumber \\
({\bf 8}',1/2)\phantom{xi}&=&(\eta _1 \bar{K})_{1/2}
\end{eqnarray}

\subsection{Two-meson "$PV$" states in definite SU(3) representations}

The labels $P$ and $V$ ($\pi$, $\rho $ etc.) 
denote two different types of resonances of appropriate flavour.

9.4.1 Intermediate states in $B^+$ decays

\noindent
a) Strangeness $S=0$
\begin{eqnarray}
({\bf 27},2)\phantom{xi}
&=&(\pi ^+ \rho ^0 +\pi ^0 \rho ^+)/\sqrt{2}\nonumber\\
\left[
\begin{array}{r}
({\bf 27},1) \\ ({\bf 8},1)_s
\end{array}
\right]
&=&
\frac{1}{\sqrt{5}}
\left[
\begin{array}{rr}
\sqrt{3} & -\sqrt{2} \\
\sqrt{2} & \sqrt{3}
\end{array}
\right]
\left[
\begin{array}{c}
(\pi ^+ \omega _8 + \eta _8 \rho ^+ )/\sqrt{2}\\
(K^+ \bar{K} ^{*0} + \bar{K} ^0 K^{*+})/\sqrt{2}
\end{array}
\right]
\nonumber\\
({\bf 8},1)_a \phantom{xi}
&=&(\sqrt{2}(\pi ^+ \rho ^0-\pi ^0 \rho ^+)-
(K^+ \bar{K} ^{*0}-\bar{K} ^0 K^{*+}))/\sqrt{6}\nonumber\\
({\bf 8},1)_{81}\phantom{x}&=&\pi ^+ \omega _1\nonumber\\
({\bf 8},1)_{18}\phantom{i.}&=&\eta _1 \rho ^+
\end{eqnarray}

\noindent
b) Strangeness $S=+1$
\begin{eqnarray}
({\bf 27},3/2)\phantom{xi}&=&(K^0\rho ^+ + \pi ^+K^{*0}+
\sqrt{2}(K^+\rho ^0+\pi ^0K^{*+}))/\sqrt{6}\nonumber\\
\left[
\begin{array}{r}
({\bf 27},1/2) \\ ({\bf 8},1/2)_s
\end{array}
\right]
&=&
\frac{1}{\sqrt{10}}
\left[
\begin{array}{rr}
1 & 3 \\
3 & -1
\end{array}
\right]
\left[
\begin{array}{c}
(K^+ \rho ^0 + \pi ^0 K^{*+} -
\sqrt{2}(K^0 \rho ^+ + \pi ^+ K^{*0}))/\sqrt{6}\\
(K^+ \omega _8 + \eta _8 K^{*+})/\sqrt{2}
\end{array}
\right]\nonumber\\
({\bf 8},1/2)_a \phantom{xi}
&=&-\frac{1}{2\sqrt{3}}(\pi ^0 K^{*+} - K^+ \rho ^0)+
\frac{1}{\sqrt{6}}(\pi ^+ K^{*0} - K^0 \rho ^+)-
\frac{1}{2}(\eta _8 K^{*+} - K^+ \omega _8)\nonumber\\
({\bf 8},1/2)_{81} \phantom{x}&=& K^+ \omega _1\nonumber\\
({\bf 8},1/2)_{18} \phantom{x}&=& \eta _1 K^{*+}
\end{eqnarray}

9.4.2 Intermediate states in $B^0_d$, $B^0_s$ decays

\noindent
a) Strangeness $S=0$
\begin{eqnarray}
({\bf 27},2) \phantom{xi}
&=& (\pi ^+ \rho ^- + \pi ^- \rho ^+ + 2 \pi ^0 \rho ^0)/\sqrt{6}
\nonumber\\
\left[
\begin{array}{r}
({\bf 27},1) \\ ({\bf 8},1)_s
\end{array}
\right]
&=&
\frac{1}{\sqrt{5}}
\left[
\begin{array}{rr}
\sqrt{3} & -\sqrt{2} \\
\sqrt{2} & \sqrt{3}
\end{array}
\right]
\left[
\begin{array}{c}
(\pi ^0 \omega _8 + \eta _8 \rho ^0 )/\sqrt{2}\\
(K^0 \bar{K} ^{*0} + K^+ K^{*-} +\bar{K} ^0 K^{*0} + K^- K^{*+})/2
\end{array}
\right]
\nonumber\\
\left[
\begin{array}{r}
({\bf 27},0) \\ ({\bf 8},0)_s \\ ({\bf 1},0)
\end{array}
\right]
&=&
\left[
\begin{array}{rrr}
\frac{1}{2\sqrt{10}}&\sqrt{\frac{3}{10}}&-\frac{3\sqrt{3}}{2\sqrt{10}}\\
\sqrt{\frac{3}{5}}&\frac{1}{\sqrt{5}}&\frac{1}{\sqrt{5}}\\
-\frac{\sqrt{3}}{2\sqrt{2}}&\frac{1}{\sqrt{2}}&\frac{1}{2\sqrt{2}}
\end{array}
\right]
\left[
\begin{array}{c}
(\pi ^+ \rho ^- + \pi ^- \rho ^+ - \pi ^0 \rho ^0)/\sqrt{3}\\
(K^+ K^{*-} +K^- K^{*+} -K^0 \bar{K} ^{*0} - \bar{K} ^0 K^{*0})/2\\
\eta _8 \omega _8
\end{array}
\right]\nonumber\\
({\bf 8},1)_a \phantom{xi}
&=& (2(\pi ^+ \rho ^- -\pi ^- \rho ^+)-
(K^+ K^{*-} + K^0 \bar{K} ^{*0}- K^- \bar{K} ^{*+} - 
\bar{K} ^0 K^{*0}))/(2\sqrt{3})\nonumber\\
({\bf 8},0)_a \phantom{xi}
&=& (K^+ K^{*-}-K^0 \bar{K} ^{*0} - K^- K^{*+} + 
\bar{K} ^0 K^{*0})/2\nonumber\\
({\bf 8},1)_{81} \phantom{x} &=& \pi ^0 \omega _1 \nonumber\\
({\bf 8},0)_{81} \phantom{x}&=& -\eta _8 \omega _1 \nonumber \\
({\bf 8},1)_{18} \phantom{x}&=& \eta _1 \rho _0 \nonumber \\
({\bf 8},0)_{18} \phantom{x}&=& -\eta _1 \omega _8 \nonumber \\
({\bf 1},0)_{11} \phantom{x}&=& \eta _1 \omega _1 
\end{eqnarray}

\noindent
b) Strangeness $S=+1$
\begin{eqnarray}
({\bf 27},3/2)\phantom{xi} &=& (K^+ \rho ^- + \pi ^- K^{*+} +
\sqrt{2}(K^0 \rho ^0 + \pi ^0 K^{*0}))/\sqrt{6}\nonumber\\
\left[
\begin{array}{r}
({\bf 27},1/2) \\ ({\bf 8},1/2)_s
\end{array}
\right]
&=&
\frac{1}{\sqrt{10}}
\left[
\begin{array}{rr}
1 & 3 \\
3 & -1
\end{array}
\right]
\left[
\begin{array}{c}
(\sqrt{2}(\pi ^- K^{*+} + K^+ \rho ^-)-
(\pi ^0 K^{*0} + K^0 \rho ^0))/\sqrt{6}\\
(\eta _8 K^{*0} + K^0 \omega _8)/\sqrt{2}
\end{array}
\right]
\nonumber \\
({\bf 8},1/2)_a\phantom{xi}
&=&(\sqrt{2}(K^+ \rho ^- -\pi ^- K^{*+})-
(K^0 \rho ^0 - \pi ^0 K^{*0}))/\sqrt{12}+(K^0 \omega _8 -\eta _8 K^{*0})/2
\nonumber\\
({\bf 8},1/2)_{81}\phantom{x}&=& K^0 \omega _1 \nonumber \\
({\bf 8},1/2)_{18}\phantom{x}&=& \eta _1 K^{*0}
\end{eqnarray}

\noindent
c) Strangeness $S=-1$
\begin{eqnarray}
({\bf 27},3/2)\phantom{xi}&=&(\pi ^+ K^{*-} + K^- \rho ^+ +
\sqrt{2}(\pi ^0 K^{*0}+\bar{K} ^0 \rho ^0))/\sqrt{6}\nonumber\\
\left[
\begin{array}{r}
({\bf 27},1/2) \\ ({\bf 8},1/2)_s
\end{array}
\right]
&=&
\frac{1}{\sqrt{10}}
\left[
\begin{array}{rr}
1 & 3 \\
3 & -1
\end{array}
\right]
\left[
\begin{array}{c}
(\sqrt{2}(\pi ^+ K^{*-} + K^-\rho ^+)-
(\pi ^0 \bar{K} ^{*0} +\bar{K} ^0 \rho ^0))/\sqrt{6} \\ 
(\eta _8 \bar{K} ^{*0}+\bar{K} ^0 \omega _8)/\sqrt{2}
\end{array}
\right]
\nonumber \\
({\bf 8},1/2)_a\phantom{x}&=&(\sqrt{2}(\pi ^+ K^{*-} - K^- \rho ^+)-
(\pi ^0 \bar{K}^{*0}-\bar{K} ^0 \rho ^0))/\sqrt{12}+
(\eta _8 \bar{K} ^{*0} - \bar{K} ^0 \omega _8)/2\nonumber\\
({\bf 8},1/2)_{81}\phantom{x}&=&\bar{K} ^0 \omega _1\nonumber\\
({\bf 8},1/2)_{18}\phantom{x}&=&\eta _1 \bar{K} ^{*0}
\end{eqnarray}

\vfill
FIGURE CAPTIONS

Fig.1 Quark-line diagrams for $B$ decays

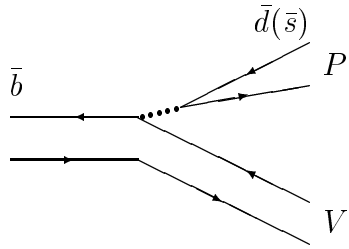
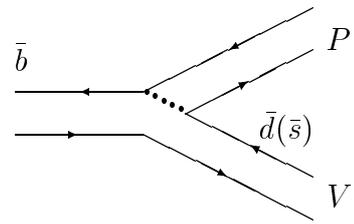
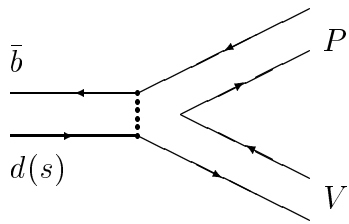
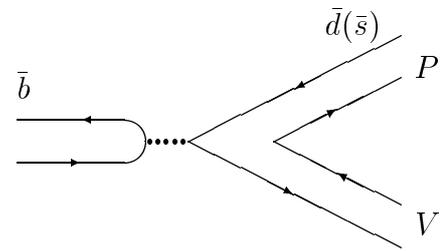
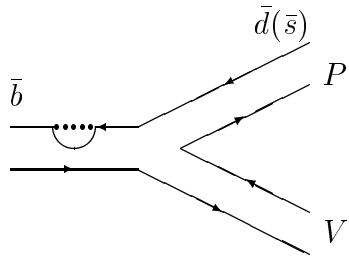
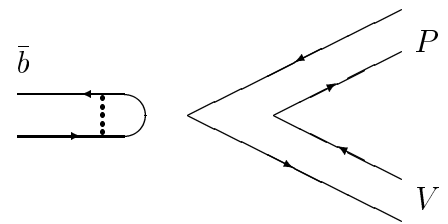
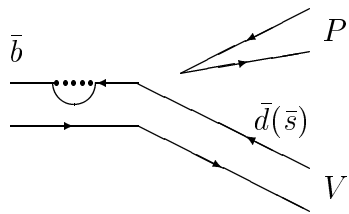
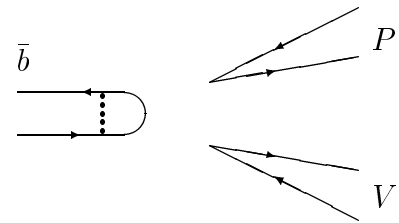
\end{document}